\documentclass[prd,showpacs]{revtex4}
\usepackage{epsfig}
\usepackage{amsmath}
\usepackage{amssymb}

\begin{document}

\title{ Study of $B_{s}\to \pi \rho$  decays
in the perturbative QCD approach}

\author{Xian-Qiao Yu\footnote{yuxq@mail.ihep.ac.cn},
 Ying Li\footnote{liying@mail.ihep.ac.cn}}

\affiliation
 {\it \small Institute of High
Energy Physics, P.O.Box 918(4), Beijing 100049, China;}
\affiliation{\it \small  Graduate School of the Chinese Academy of
Sciences, Beijing 100049, China}

\author{Cai-Dian L\"u}

\affiliation{\it \small  CCAST (World Laboratory), P.O. Box 8730,
Beijing 100080, China;}
 \affiliation{\it\small  Institute of High
Energy Physics, P.O.Box 918(4), Beijing 100049,
China\footnote{Mailing address} }

\date{\today}

\begin{abstract}
  In this note, we calculate the
branching ratio and $CP$ asymmetry parameters of
$B_s\rightarrow\pi\rho$ in the framework of perturbative QCD
approach based on $k_T$ factorization. This decay can occur only
via annihilation diagrams in the Standard Model. We find that (a)
the charge averaged
$Br(B_s\rightarrow\pi^{+}\rho^{-}+\pi^{-}\rho^{+})$ is about
$(9-12)\times10^{-7}$; $Br(B_s\rightarrow\pi^{0}\rho^{0})
\simeq4\times10^{-7}$; and (b) there are sizable $CP$ asymmetries
in the processes, which can be tested in the near future Large
Hadron Collider beauty experiments (LHC-b) at CERN.
\end{abstract}

 \maketitle

The $B$ meson rare decays provide a good place for testing the
Standard Model (SM), studying $CP$ violation and looking for
possible new physics beyond the SM. A lot of theoretical studies
have been done  in recent years, which are strongly supported by
the running $B$ factories in KEK and Stanford Linear Accelerator
Center (SLAC).   Looking forward to the future CERN Large Hadron
Collider beauty experiments (LHC-b), a large number of $B_{s}$ and
$B_{c}$ mesons can also be produced. So the studies of $B_{s}$
meson rare decays are necessary in the next a few years.

In this paper, we study the rare decays $B_{s}\to \pi \rho$ in
Perturbative QCD approach (PQCD) \cite{LY}. Comparing with QCD
factorization approach \cite{bbns}, PQCD approach can make a
reliable calculation for pure annihilation diagrams in $k_T$
factorization. The endpoint singularity occurred in QCD
factorization approach can be cured here by the Sudakov factor
from resummation of double logarithms.

 In PQCD approach, the decay amplitude can be written as:
\begin{equation}
 \mbox{Amplitude}
\sim \int\!\! d^4k_1 d^4k_2 d^4k_3\ \mathrm{Tr} \bigl[ C(t)
\Phi_{B_s}(k_1) \Phi_{\pi}(k_2) \Phi_\rho(k_3) H(k_1,k_2,k_3, t)
\bigr]e^{-S(t)}. \label{eq:convolution1}
\end{equation}
In our following calculations, the Wilson coefficient $C(t)$,
Sudakov factor $S_{i}(t)(i=B_s, \pi, \rho)$ and the
non-perturbative but universal wave function $\Phi_{i}$ can be
found in the Refs. \cite{LUY,YLL,BFB,BBKT,LYE}. The  hard part $H$
are channel dependent but fortunately perturbative calculable,
which will be shown below.

\begin{figure}[htb]
\begin{center}
\includegraphics[scale=0.70]{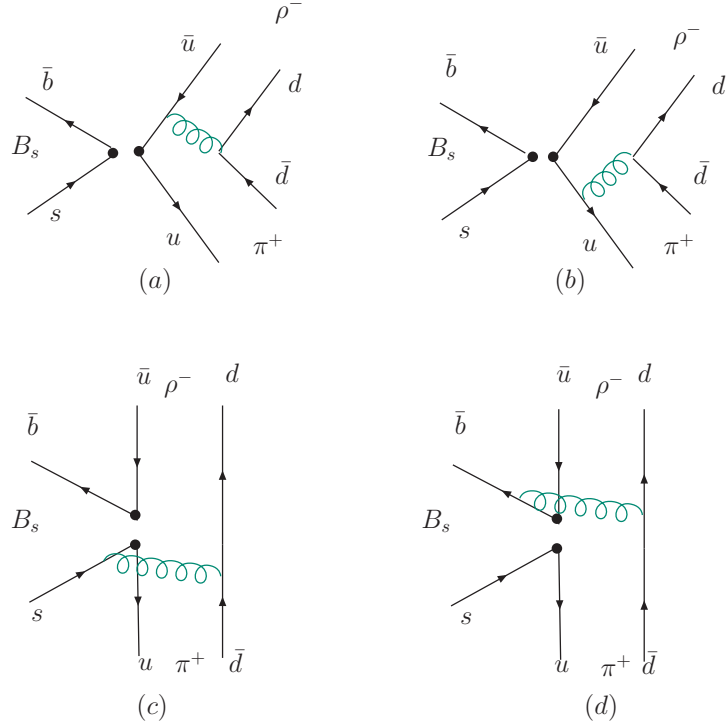}
\caption{The lowest order diagrams for
$B_{s}\rightarrow\pi^{+}\rho^{-}$ decay} \label{figure:fig1}
\end{center}
\end{figure}

Like the $B_s\rightarrow\pi^{+}\pi^{-}$ decay \cite{LLXY}, the
$B_s\rightarrow\pi \rho$ decays are pure annihilation type rare
decays, which are difficult to be calculated in method other than
PQCD approach. Fig.~\ref{figure:fig1} shows the lowest order
Feynman diagrams to be calculated in PQCD approach where the big
dots denote the quark currents in four quark operators for
$B_s\rightarrow\pi^{+}\rho^{-}$ according to the effective
hamiltonian of $b$ quark decay \cite{BBL}. First for the usual
factorizable diagram (a) and (b), the sum of (V-A)(V-A) current
contributions is given by
\begin{multline}
M_a[C]  = 16\pi C_F M_B^2 \int_0^1\!\!\!
 dx_2 dx_3
  \int_0^\infty\!\!\!\!\! b_2 db_2\, b_3 db_3\
  \bigl\{[
 x_{3}\phi_{\pi}^A(x_2)\phi_{\rho}(x_3)+2r_{\pi}r_{\rho}(1+x_{3})\phi_{\pi}^{P}(x_2)\phi_{\rho}^{s}(x_3)\\
 +2r_{\pi}r_{\rho}(x_{3}-1)\phi_{\pi}^{P}(x_2)\phi_{\rho}^{t}(x_3)]
 \alpha_{s}(t_{a}^{1})h_{a}(x_{2},x_3,b_2,b_3)
 \exp[-S_\pi(t_{a}^{1})-S_\rho(t_{a}^{1})]C(t_{a}^{1})\\
-[x_{2}\phi_{\pi}^A(x_2)\phi_{\rho}(x_3)+2r_{\pi}r_{\rho}(1+x_{2})\phi_{\pi}^{P}(x_2)\phi_{\rho}^{s}(x_3)
 +2r_{\pi}r_{\rho}(x_{2}-1)\phi_{\pi}^{T}(x_2)\phi_{\rho}^{s}(x_3)]\\
 \alpha_{s}(t_{a}^{2})
h_{a}(x_{3},x_2,b_3,b_2)\exp[-S_\pi(t_{a}^{2})-S_\rho(t_{a}^{2})]C(t_{a}^{2})
\bigr\},\label{ma}
\end{multline}
where $r_{\pi}=m_{0\pi}/m_B=m_{\pi}^2/[m_B(m_u+m_d)]$,
 $r_\rho=m_{\rho}/m_B$. $C_F=4/3$ is the group factor of the $SU(3)_{c}$ gauge
 group. $\phi_i$ is light cone distribution amplitude, which describes the momentum distribution
  of the meson wave function.
  The function
 \begin{align}
& h_{a}(x_2,x_3,b_2,b_3) = S_{t}(x_3)\frac{\pi i}{2}H_{0}^{(1)}
(M_{B}\sqrt{x_2x_3}b_{2})\nonumber \\
&\times \bigl[\theta(b_2-b_3)J_{0}(M_{B}\sqrt{x_3}b_3)\frac{\pi
i}{2}
  H_{0}^{(1)}(M_{B}\sqrt{x_3}b_2)+(b_2
\leftrightarrow b_3)\bigr]
 \label{eq:propagator4}
\end{align}
comes from the Fourier transformation of propagators of virtual
quark and gluon in the hard part calculations. $S_{t}(x)$ results
from the threshold resummation of QCD radiative corrections to
hard amplitudes \cite{KLS}. $H_{0}^{1}(z)=J_{0}(z)+iY_0(z)$,
$J_{0}$ and $Y_{0}$ are Bessel functions. The sum of (V-A)(V+A)
current contributions of diagrams (a) and (b) is $-M_a[C]$.

For the non-factorizable annihilation diagrams (c) and (d), all
three meson wave functions are involved. The (V-A)(V-A) operator's
contribution is:
\begin{multline}
M_c[C]  =\frac{-1}{\sqrt{2N_{c}}}64\pi C_F M_B^2 \int_0^1\!\!\!
 dx_1 dx_2dx_3
  \int_0^\infty\!\!\!\!\! b_1 db_1\ b_2 db_2\
 \phi_B(x_1,b_1) \\
 \times
 \bigl\{[-x_{2}\phi_{\rho}(x_3)\phi_{\pi}^A(x_2)
 +r_{\rho}r_{\pi}(x_3-x_2)\phi_{\rho}^{t}(x_3)\phi_{\pi}^P(x_2)
 -r_{\rho}r_{\pi}(x_2+x_3)\phi_{\rho}^{s}(x_3)\phi_{\pi}^P(x_2)
 \\
 -r_\pi r_\rho
 (x_2+x_3)\phi_{\rho}^{t}(x_3)\phi_{\pi}^T(x_2)+r_\pi
 r_\rho(x_3-x_2)\phi_{\rho}^{s}(x_3)\phi_{\pi}^T(x_2)]
 \\ C(t_c^1)
  \alpha_{s}(t_{c}^{1})h_{c}^{(1)}(x_{1},x_2,x_3,b_1,b_2)\exp[-S_B(t_{c}^{1})
 -S_{\pi}(t_{c}^{1})-S_\rho(t_{c}^{1})]
  -[-x_3\phi_{\rho}(x_3)\phi_{\pi}^A(x_2)\\
  +r_{\rho}r_{\pi}(x_2-x_3)\phi_{\rho}^{t}(x_3)\phi_{\pi}^P(x_2)
 -r_{\rho}r_{\pi}(2+x_2+x_3)\phi_{\rho}^{s}(x_3)\phi_{\pi}^P(x_2)
  \\
  +r_\pi r_\rho(2-x_2-x_3)\phi_{\rho}^{t}(x_3)\phi_{\pi}^T(x_2)
 +r_\pi r_\rho(x_2-x_3)\phi_{\rho}^{s}(x_3)\phi_{\pi}^T(x_2)]
  \\C(t_c^2)
 \alpha_{s}(t_{c}^{2})h_{c}^{(2)}(x_{1},x_2,x_3,b_1,b_2)\exp[-S_B(t_{c}^{2})
 -S_{\pi}(t_{c}^{2})-S_\rho(t_{c}^{2})]
 \bigr\}.\label{mc}
\end{multline}
For the penguin operators, there are also $(V-A)(V+A)$ type
operators, whose  contribution is given as
\begin{multline}
M_c^P[C]  =\frac{1}{\sqrt{2N_{c}}}64\pi C_F M_B^2 \int_0^1\!\!\!
 dx_1 dx_2dx_3
  \int_0^\infty\!\!\!\!\! b_1 db_1\ b_2 db_2\
 \phi_B(x_1,b_1) \\
 \times
 \bigl\{[-x_3\phi_{\rho}(x_3)\phi_{\pi}^A(x_2)
 -r_{\rho}r_{\pi}(x_3-x_2)\phi_{\rho}^{t}(x_3)\phi_{\pi}^P(x_2)
 -r_{\rho}r_{\pi}(x_2+x_3)\phi_{\rho}^{s}(x_3)\phi_{\pi}^P(x_2)
 \\
 -r_\pi r_\rho
 (x_2+x_3)\phi_{\rho}^{t}(x_3)\phi_{\pi}^T(x_2)-r_\pi
 r_\rho(x_3-x_2)\phi_{\rho}^{s}(x_3)\phi_{\pi}^T(x_2)]
 \\ C(t_c^1)
  \alpha_{s}(t_{c}^{1})h_{c}^{(1)}(x_{1},x_2,x_3,b_1,b_2)\exp[-S_B(t_{c}^{1})
 -S_{\pi}(t_{c}^{1})-S_\rho(t_{c}^{1})]
  -[-x_{2}\phi_{\rho}(x_3)\phi_{\pi}^A(x_2)\\
  -r_{\rho}r_{\pi}(x_2-x_3)\phi_{\rho}^{t}(x_3)\phi_{\pi}^P(x_2)
 -r_{\rho}r_{\pi}(2+x_2+x_3)\phi_{\rho}^{s}(x_3)\phi_{\pi}^P(x_2)
  \\
  +r_\pi r_\rho(2-x_2-x_3)\phi_{\rho}^{t}(x_3)\phi_{\pi}^T(x_2)
 -r_\pi r_\rho(x_2-x_3)\phi_{\rho}^{s}(x_3)\phi_{\pi}^T(x_2)]
  \\ C(t_c^2)
 \alpha_{s}(t_{c}^{2})h_{c}^{(2)}(x_{1},x_2,x_3,b_1,b_2)\exp[-S_B(t_{c}^{2})
 -S_{\pi}(t_{c}^{2})-S_\rho(t_{c}^{2})]
 \bigr\},\label{mcp}
\end{multline}
where
\begin{align}
& h^{(j)}_c(x_1,x_2,x_3,b_1,b_2) = \nonumber \\
& \biggl\{\theta(b_2-b_1) \frac{\pi i}{2}
\mathrm{H}_0^{(1)}(M_B\sqrt{x_2x_3}\, b_2)
 \mathrm{J}_0(M_B\sqrt{x_2x_3}\, b_1)
\nonumber \\
& \qquad\qquad\qquad\qquad + (b_1 \leftrightarrow b_2) \biggr\}
 \times\left(
\begin{matrix}
 \mathrm{K}_0(M_B F_{(j)} b_1), & \text{for}\quad F^2_{(j)}>0 \\
 \frac{\pi i}{2} \mathrm{H}_0^{(1)}(M_B\sqrt{|F^2_{(j)}|}\ b_1), &
 \text{for}\quad F^2_{(j)}<0
\end{matrix}\right),
\label{eq:propagator3}
\end{align}
$K_{0}$ is modified Bessel function and $F_{(j)}$'s are defined by
\begin{equation}
 F^2_{(1)} =x_1x_2-x_2x_3;\
F^2_{(2)} =x_1 +x_2+x_3-x_1x_2-x_2x_3.
\end{equation}

The hard scale $t_{i}'s$ in Eqs.(\ref{ma}, \ref{mc}, \ref{mcp})
are chosen as the largest energy scale appearing in each diagram
to kill the large logarithmic corrections:
\begin{eqnarray}
\nonumber t_a^{1} &=& \mathrm{max}(M_B \sqrt{x_{3}},
 1/b_2,1/b_3), \\
\nonumber t_a^{2} &=& \mathrm{max}(M_B \sqrt{x_{2}},
 1/b_2,1/b_3), \\
\nonumber t_c^{1} &=& \mathrm{max}(M_B
\sqrt{|F^2_{(1)}|},M_B\sqrt{x_{2}x_{3}},
 1/b_1,1/b_2), \\
 \nonumber t_c^{2} &=& \mathrm{max}(M_B
\sqrt{|F^2_{(2)}|},M_B\sqrt{x_{2}x_{3}},
 1/b_1,1/b_2).
\end{eqnarray}

The total decay amplitude is then
\begin{align}
{ \cal A} (B_s \to \pi^+\rho^-)
=f_{B}M_{a}\left[V_{ub}^{*}V_{us}(C_{1}+\frac{1}{3}C_{2})-V_{tb}^{*}V_{ts}(
  2C_{3}+\frac{2}{3}C_{4}-2C_5-\frac{2}{3}C_6-\frac{1}{2}C_7-\frac{1}{6}C_{8}+\frac{1}
 {2}C_9+\frac{1}{6}C_{10})\right]
  \nonumber\\
+M_{c}\left[V_{ub}^{*}V_{us}C_{2}-V_{tb}^{*}V_{ts}(2C_{4}+\frac{1}{2}C_{10})
  \right]
  -V_{tb}^{*}V_{ts}M_{c}^{P}\left(2C_{6}+\frac{1}{2}C_{8}\right),\hspace*{5cm}.
\end{align}
and the decay width is expressed as

\begin{equation}
 \Gamma(B_{s} \to \pi^+ \rho^{-}) = \frac{G_F^2 M_B^3}{128\pi}
\left| { \cal A} (B_{s}\rightarrow\pi^{+}\rho^{-})\right|^2.
\label{eq:width3}
\end{equation}

The most important contribution here is the factorizable  penguin
diagram, which is CKM enhanced. If we exchange the $\pi$ and
$\rho$ in Fig. 1, by the same method, we can compute the
$B_s\rightarrow\pi^{-}\rho^{+}$ decay. The expressions are
similar. The decay amplitude for $B_s \to \pi^0 \rho^0$ is
\begin{equation}
{ \cal A} (B_s \to \pi^0\rho^0)= { \cal A} (B_s \to \pi^+\rho^-)+{
\cal A} (B_s \to \pi^-\rho^+),
\end{equation}
and the decay width can be written as

\begin{equation}
 \Gamma(B_{s} \to \pi^0 \rho^{0}) = \frac{G_F^2 M_B^3}{512\pi}
\left| { \cal A} (B_s \to \pi^0\rho^0) \right|^2.
\label{eq:width4}
\end{equation}

In the following, we first give the branching ratios of
$B_s\rightarrow\pi\rho$. Just as in Ref. \cite{LLXY}, we leave the
CKM angle $\gamma$ as a free parameter in our numerical
calculations. Because there are four decay channels:
$B_s/\bar{B}_s\rightarrow\pi^{+}\rho^{-}$,
$B_s/\bar{B}_s\rightarrow\pi^{-}\rho^{+}$, it is not possible  to
distinguish the initial state by detecting the final states.  We
average the sum of $B_s/\bar{B}_s\rightarrow\pi^{+}\rho^{-}$ as
one channel, and $B_s/\bar{B}_s\rightarrow\pi^{-}\rho^{+}$ as
another, which is distinguishable by experiments. Using the same
parameters as refs. \cite{YLL,LYE}, we get
\begin{align}
\mathbf{Br}(B_s/\bar{B}_s\rightarrow\pi^{+}\rho^{-})=(5.1^{+0.8}_{-0.5})\times10^{-7},\nonumber \\
\mathbf{Br}(B_s/\bar{B}_s\rightarrow\pi^{-}\rho^{+})=(5.4^{+0.5}_{-0.8})\times10^{-7},\nonumber \\
\mathbf{Br}(B_s/\bar{B}_s\rightarrow\pi^{0}\rho^{0})=(4.2^{+0.6}_{-0.7})\times10^{-7},
\end{align}
where all channels are averaged for $B_s$ and $\bar B_s$.

In Ref. \cite{BN}, Beneke $et\, al$ have estimated the branching
ratio of $\bar{B}_s\rightarrow\pi^{+}\rho^{-}$ in the QCD
factorization approach. Weak annihilation diagrams are power
suppressed in the heavy quark limit and, in general, not
calculable in QCD factorization approach. In order to avoid the
end-poind singularities, they introduced phenomenological
parameters to replace the divergent integral. With those
parameters they estimated that the branching ratio of
$\bar{B}_s\rightarrow\pi^{+}\rho^{-}$ is
$(0.03-0.14)\times10^{-7}$. In PQCD approach, the annihilation
amplitude is calculable. In the rest frame of the $B_{s}$ meson,
the d or $\bar{d}$ quark included in $\rho$ or $\pi$ has momentum
${\cal O}(M_B/4)$, and the gluon producing them has momentum
$q^{2}={\cal O}(M_B^{2}/4)$. This is a hard gluon,
 the PQCD can be safely used because of asymptotic freedom of QCD
\cite{GWP}. We have tested that the bulk of the result comes from
the region with $\alpha_{s}/\pi<0.2$, where a figure was show in
Ref. \cite{LLX}. Our predicted result is larger than their
estimation, which can be tested by the future experiments.

\begin{figure}[htb]
  \begin{center}
  \epsfig{file=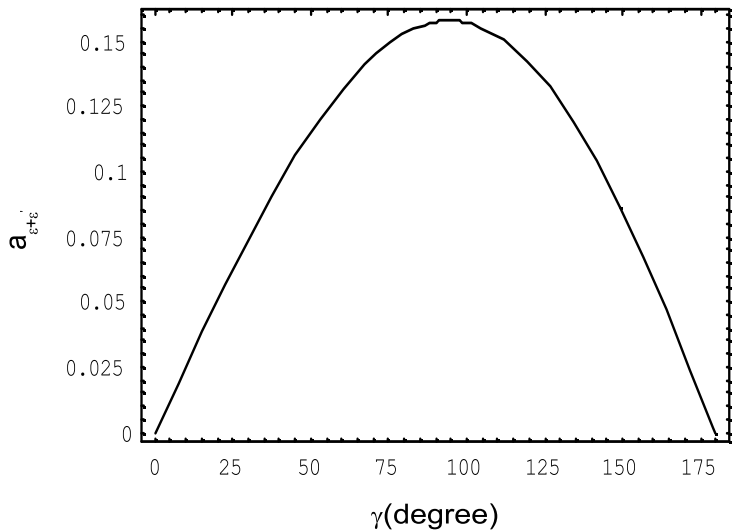,width=300pt,height=220pt}
   \end{center}
   \caption{Mixing $CP$ violation parameter of $B_{s} \to \pi^{0}
   \rho^{0}$ decay as a function of CKM angle $\gamma$}
   \label{figure:Fig2}
  \end{figure}

Using the same definition in Ref.~\cite{LLXY}, we study the $CP$
violation parameters $A^{dir}_{CP}$ and
$a_{\epsilon+\epsilon^{'}}$ in the process of
$B_s\rightarrow\pi^{0}\rho^{0}$ decay.    We find the direct $CP$
violation parameter $A^{dir}_{CP}(B_s\rightarrow\pi^{0}\rho^{0})$
is about $2\%$ when $\gamma$ is near $100^{\circ}$, the small
direct $CP$ asymmetry is also a result of small tree level
contribution. The mixing $CP$ violation parameter
$a_{\epsilon+\epsilon^{'}}(B_s\rightarrow\pi^{0}\rho^{0})$ is
large, whose peak is close to $16\%$ (see Fig.~\ref{figure:Fig2}).

The $CP$ violations of
$B_s/\bar{B}_s\rightarrow\pi^{\pm}\rho^{\mp}$ are very
complicated. There are four decay amplitudes, which can be
expressed as:
\begin{align}
g=\langle\pi^{+}\rho^{-}|H_{eff}|B_s\rangle;\hspace{1cm}
h=\langle\pi^{+}\rho^{-}|H_{eff}|\bar{B}_s\rangle;\nonumber \\
\bar{g}=\langle\pi^{-}\rho^{+}|H_{eff}|\bar{B}_s\rangle;\hspace{1cm}
\bar{h}=\langle\pi^{-}\rho^{+}|H_{eff}|B_s\rangle.
\end{align}

We introduce four parameters to describe the $CP$ asymmetries in
the processes, which are given by \cite{G}
\begin{align}
C_{\pi\rho}=\frac{1}{2}(a_{\epsilon^{'}}+a_{\bar{\epsilon}^{'}}),\hspace{1.5cm}
\Delta C_{\pi\rho}=\frac{1}{2}(a_{\epsilon^{'}}-a_{\bar{\epsilon}^{'}}),\nonumber \\
S_{\pi\rho}=\frac{1}{2}(a_{\epsilon+\epsilon^{'}}+a_{\epsilon+\bar{\epsilon}^{'}}),\hspace{1cm}
\Delta
S_{\pi\rho}=\frac{1}{2}(a_{\epsilon+\epsilon^{'}}-a_{\epsilon+\bar{\epsilon}^{'}}),
\end{align}
where
\begin{align}
a_{\epsilon^{'}}=\frac{|g|^{2}-|h|^{2}}{|g|^{2}+|h|^{2}},\hspace{1cm}
a_{\epsilon+\epsilon^{'}}=\frac{-2Im(h/g)}{1+|h/g|^{2}},\nonumber \\
a_{\bar{\epsilon}^{'}}=\frac{|\bar{h}|^{2}-|\bar{g}|^{2}}{|\bar{h}|^{2}+|\bar{g}|^{2}},\hspace{1cm}
a_{\epsilon+\bar{\epsilon}^{'}}=\frac{-2Im(\bar{g}/\bar{h})}{1+|\bar{g}/\bar{h}|^{2}}.
\end{align}

\begin{figure}[htb]
  \begin{center}
  \epsfig{file=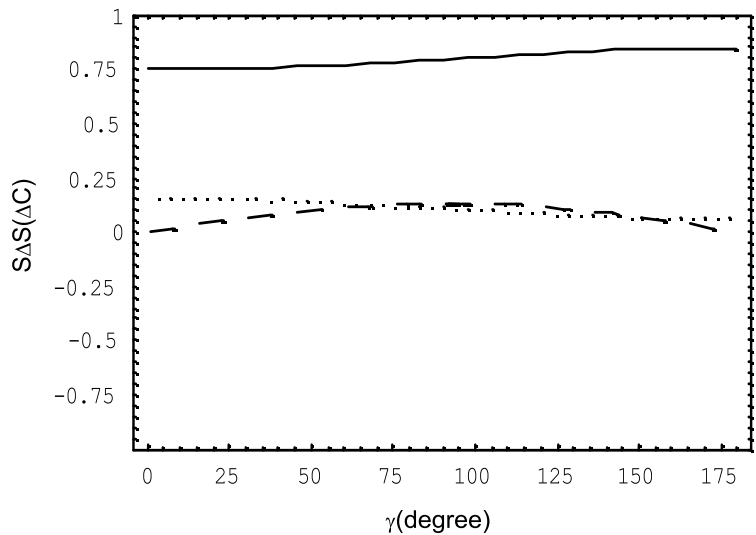,width=300pt,height=220pt}
   \end{center}
   \caption{ $CP$ violation parameters of $B_{s}^0(\bar{B}_{s}) \to \pi^{\pm}
   \rho^{\mp}$ decays: $\Delta C$ (solid line), S (dashed line) and $\Delta S$ (dotted line) as
    a function of CKM angle $\gamma$}
   \label{figure:Fig3}
  \end{figure}

We calculate the above four $CP$ asymmetry parameters and show the
CKM angle $\gamma$ dependence in Fig.~\ref{figure:Fig3}. We do not
plot the $C_{\pi\rho}$ in this figure, since its value is near
zero. The decay branching ratios depends heavily on the shape of
wave functions and decay constants etc. But the CP asymmetry
should not since the dependence will be cancelled. The direct CP
asymmetry can be affected by the power corrections and
next-to-leading order contributions easily.  We investigate the
$CP$ asymmetry parameters's dependence on the hard scale $t$ in
Eq. (\ref{eq:convolution1}), which characterize the size of
next-to-leading order contribution. The CP asymmetry numbers are
shown in Table \ref{tab1} with those uncertainties. By changing
the hard scale $t$ from $0.9t$ to $1.3t$, we find the $CP$
asymmetries of $B_s(\bar{B}_s)\rightarrow\pi^{\pm}\rho^{\mp}$
change little: for $\gamma=60^{\circ}$, the uncertainty is less
than $1\%$ for $C_{\pi\rho}$, $4\%$ for $\Delta C_{\pi\rho}$,
$3\%$ for $S_{\pi\rho}$ and $7\%$ for $\Delta S_{\pi\rho}$,
respectively. The reason is that mixing induced CP is dominant
here, but the direct CP of $B_s\rightarrow\pi^{0}\rho^{0}$ really
changed much, as shown in Table \ref{tab1}.

\begin{table}[htb]
\caption{CP asymmetry parameters using $\gamma=60^\circ$ with
uncertainties} \label{tab1}
\begin{center}
\begin{tabular}[t]{|r|c|c|c|c|c|c|}
 \hline     \hline
 Scale    & $C_{\pi\rho}$ &  $\Delta C_{\pi\rho}$  &  $S_{\pi\rho}$ &  $\Delta S_{\pi\rho} $ &
  $A^{dir}_{CP}(B_s\rightarrow\pi^{0}\rho^{0})$ &
  $a_{\epsilon+\epsilon^{'}}(B_s\rightarrow\pi^{0}\rho^{0})$ \\
\hline 0.9 $t$   & $0.1\%$ &  $73\%$  & $10\%$ &  $21\%$
& $0.6\%$ & $11\%$  \\
\hline
 $t$      & $1\%$ &  $77\%$  &  $11\%$ &  $14\%$
  & $1.6\%$ &  $13\%$ \\
\hline 1.3 $t$    & $1.8\%$ &  $79\%$  & $14\%$&  $10\%$
& $2.6\%$ &   $16\%$ \\
 \hline
\end{tabular}
\end{center}
\end{table}

In conclusion, we study the branching ratio and $CP$ asymmetries
of $B_s\rightarrow\pi\rho$ decays in PQCD approach. We find the
branching ratio of $B_s\rightarrow\pi^{+}\rho^{-}+\pi^{-}\rho^{+}$
is at order $  10^{-6}$, which is larger than QCD factorization's
estimation. We also predict $CP$ asymmetries in the process, which
may be measured in the future LHC-b experiments.

We thank M.-Z. Yang for useful discussions. This work is partly
supported by National Science Foundation of China under Grant No.
90103013, 10475085 and 10135060.

\end{document}